\documentclass[twocolumn,showpacs,aps,prb,floatfix,superscriptaddress]{revtex4}
\usepackage{amssymb}
\usepackage{amsmath}
\usepackage{epsfig}
\begin{document}
\title{
Massless Fermions in multilayer graphitic systems with misoriented layers.
}
\author{Sylvain Latil}
\affiliation{Laboratoire de Physique du Solide, 
  Facult\'es Universitaires Notre-Dame de la Paix, 
  rue de Bruxelles 61, 5000 Namur, Belgium}
\author{Vincent Meunier}
\affiliation{Oak Ridge National Laboratory, Bethel Valley Road, 
  Oak Ridge, TN, 37831-6367, USA}
\author{Luc Henrard}
\affiliation{Laboratoire de Physique du Solide, 
  Facult\'es Universitaires Notre-Dame de la Paix, 
  rue de Bruxelles 61, 5000 Namur, Belgium}

\date{\today}
\begin{abstract}
  We examine how the misorientation of a few stacked
  graphene layers affects the electronic structure of carbon
  nanosystems. 
  We present {\it ab initio} calculations on bi- and
  trilayer systems to demonstrate that the massless Fermion
  behavior typical of single layered graphene is also found in 
  incommensurate multilayered graphitic systems. 
  We also investigate the consequences of this property 
  on experimental fingerprints, 
  such as Raman spectroscopy and scanning tunneling microscopy (STM). 
  Our
  simulations reveal that STM images of turbostratic few layer
  graphite are sensitive to the layer arrangement. 
  We also predict that resonant raman signal of graphitic samples 
  are more sensitive to the orientation of the layers than to their number.
\end{abstract}
\maketitle
%
%

The electronic properties of 2D graphene and 3D graphite have been
extensively studied for more than 50
years~\cite{McClure_Slonczewski}.
It is fascinating that the extraordinary properties of carbon
nanotubes \cite{Charlier_RevModPhys_79_677_2007} have been deduced
from those of 2D graphene many years before macroscopic samples of
very thin Few Layer {Graphite} (FLG) could be obtained in the 
laboratory~\cite{Novoselov_Berger}.
A particular interest has been
recently given to Single Layer Graphene (SLG) because of their massless
Fermion behavior, the $\sqrt{B}$ dependence of the Landau
levels~\cite{Gusynin_PhysRevB_73_245411_2006,
Sadowski_PhysRevLett_97_266405_2006}, and the observation of abnormal
Quantum Hall Effect (QHE), even at room
temperature~\cite{Novoselov_and_co}.

In that context, a precise investigation of the layer to layer
interaction on the existence of massless Fermion carriers is 
of paramount importance. 
For 3D graphite, the most stable Bernal phase (AB stacking)
as well as the rhombohedral (ABC stacking) have been proven to show
complex electron and hole bands near the Fermi level rather
than linear, massless fermion ones, due to interlayer
interaction~\cite{Charlier}.
We have recently shown that regular (AB or ABC) stackings also break
the linear character of the dispersion of electronic bands for FLG
with 2 to 4 layers and that ambipolar electronic conduction could be
related to AB stacked FLG~\cite{Latil_PhysRevLett_98_036803_2006}.  

In this Letter, we present an electronic structure analysis of
misoriented (turbostratic) 2 and 3 layer FLG. A recent surface X-rays
analysis of multilayer graphene grown on SiC shows that misorientation
is plausible~\cite{Hass_condmat0702540}.  We show here that the linear
dispersion of SLG is preserved in turbostratic multilayer systems
despite the presence of adjacent layers.  It follows that massless
Fermion carriers are predicted for disoriented multilayer systems.
These findings raise the question of the interpretation of the
experimental observations performed on FLG samples and challenge the
direct relation between SLGs and Dirac massless Fermions.  More
generally, the electronic properties (and consequently the optical,
vibrational, and transport properties) of a given FLG film is found to
be controlled mainly by the (mis)orientation of the successive layers
rather than the number of layers.  In the present work , we discuss
also the implications of this possible misorientation on experimental
signatures, in particular on STM and Raman fingerprints.

\begin{figure}[b]
\epsfig{file=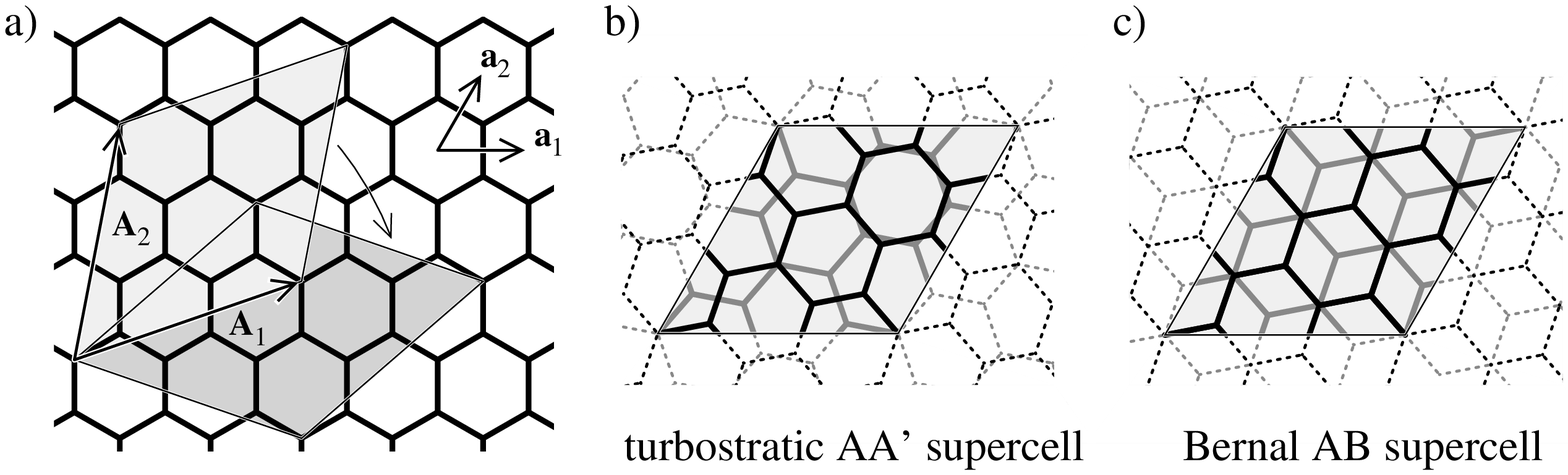, width=\linewidth, clip=}
\caption{a) The non-symmetric supercell $(2,1)$ (light grey), 
and its two primitive vectors ${\bf A}_1$ and ${\bf A}_2$. 
The $(3,\overline{1})$ cell is drawn in dark grey.
b) A quasi incommensurate bilayer AA' structure, made by 
stacking the $(2,1)$ on the $( 3,\overline{1})$ supercell.
c) An AB bilayer structure supercell.
}
\label{f:method}
\end{figure}

The main technical difficulty when modeling turbostratic structures, is
to combine their incommensurate character and the necessary finiteness
of the supercell in solid state calculations.  
An elegant solution, proposed by
Kolmogorov and {Crespi~\cite{Kolmogorov_PhysRevB_71_235415_2005}}
consists in the definition of a graphene hexagonal supercell without
mirror symmetry (except the basal plane). 
The smallest supercell of
this type contains 14 atoms, and is defined with its basis vectors
${\bf A}_1=2{\bf a}_1+{\bf a}_2$ (denoted as $(2,1)$ supercell
hereafter) and ${\bf A}_2=-{\bf a}_1+3{\bf a}_2$, as shown on
Fig.\ref{f:method}. 
The supercell $(3,\overline{1})$
possesses exactly the same basis vectors, but is rotated by the angle
cos$^{-1}(11/14)\simeq$ {38.21}$^{\circ}$. 
When stacking these two supercells, we
obtain a bilayer structure with short-range incommensurability
(Fig.\ref{f:method}).
The $(7,0)$ and $(5,3)$ supercells also present
exactly the same size but a larger number of atoms (98 C atoms per
layer), and can be assembled onto bi- or trilayer compounds
fulfilling the translational symmetry 
requirements~\cite{Note:3layers}.

In this Letter, we focus our attention on bi- and trilayer FLG
structures based on the $(2,1)$ and $(7,0)$/$(5,3)$ supercells. 
Our
aim is to highlight significant differences in the band structures of
commensurate and incommensurate graphitic systems. All the
calculations were performed with density functional theory (DFT),
within the local density approximation (LDA)
scheme~\cite{Goedecker_PhysRevB_54_1703_1996}, and using
norm-conserving
pseudopotentials~\cite{Troullier_PhysRevB_43_8861_1991}.  According to
the number of atoms involved in the supercell, the eigen-problem was
expanded on plane waves (PWs, using the code {\sc abinit}
\cite{ABINIT}) or localized pseudoatomic orbitals (PAOs, using the
code {\sc siesta}~\cite{Soler_JPhysCondensMatter_14_2745_2002}).  In
the first case, plane waves cutoff energy was set to 35 Hartree. In
the second case, the basis was composed of atom-centered
double-$\zeta$ functions.

\begin{figure}[t]
\epsfig{file=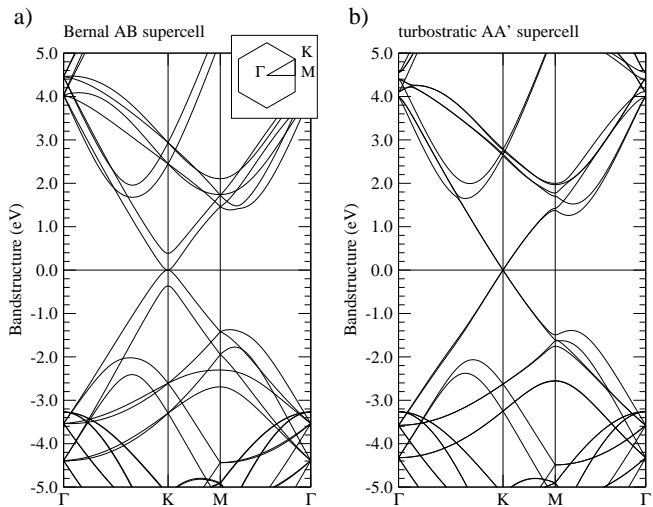, 
  width=\linewidth, 
  clip=}
\caption{ The 2D electronic band structures of Bernal AB (a) and
turbostratic AA' (b) bilayer graphene. 
Both are based on the supercell represented in Fig. 1.
The hexagonal Brillouin zone of the system is shown in the inset.}
\label{f:bsI}
\end{figure}

We carried out a structural optimization of each bilayer $(2,1)$-based 
supercell (PW calculations with tolerance for the forces set to
10$^{-6}$ Hartree/Bohr and a 12$\times$12$\times$1 grid sampling of 
the Brillouin zone), prior to the analysis of their electronic structure.  
No significant atomic rearrangement is observed within a
plane, the only structural reorganization taking place in the average
inter-layer distance (3.33~{\AA} for the AB, and 3.42~{\AA} for the
turbostratic stacked bilayer).  
The corresponding band structures of
the two (2,1)-based bilayers are plotted on Fig.~\ref{f:bsI}. We have
carefully verified that PAO basis yields identical results.

Fig.~\ref{f:bsI}a {reproduces} the well-known results for AB
bilayer~\cite{Latil_PhysRevLett_98_036803_2006,Graf_Nanolett_7_238_2007}
within a supercell $\sqrt{7}\times\sqrt{7}$ larger then the primitive
cell and is {displayed} for comparison. As in previously published
calculations, we notice a $\sim0.8$ eV splitting of the electronic
bands, related to the layer-layer interactions in the AB geometry. The
bilayer graphene with a 38.21$^{\circ}$ misorientation {(called AA'
  hereafter)} presents a totally different signature near the Fermi
level (Fig.\ref{f:bsI}b). Surprisingly, the band structure here is
found to be similar to the one of SLG but doubly
degenerate. The significance of this result is striking, since a
massless Fermion character is found for misoriented bilayer systems
with a Fermi velocity of $\sim$9.6$\times$10$^6$~m s$^{-1}$,
{strictly} identical to the value deduced for SLG within the same
formalism, and similar to experimental
values~\cite{Novoselov_and_co,Lukyanchuk_Zhou}.  We have checked that
this unexpected behavior is also found in the turbostratic
$(5,3)$on$(8,\overline{3})$ bilayer that presents a misorientation
angle of 43.57$^{\circ}$.  The apparent absence of interaction
between layers at the Fermi level can be related to the loss of short
range corrugation between successive layers.

\begin{figure}[b]
\epsfig{file=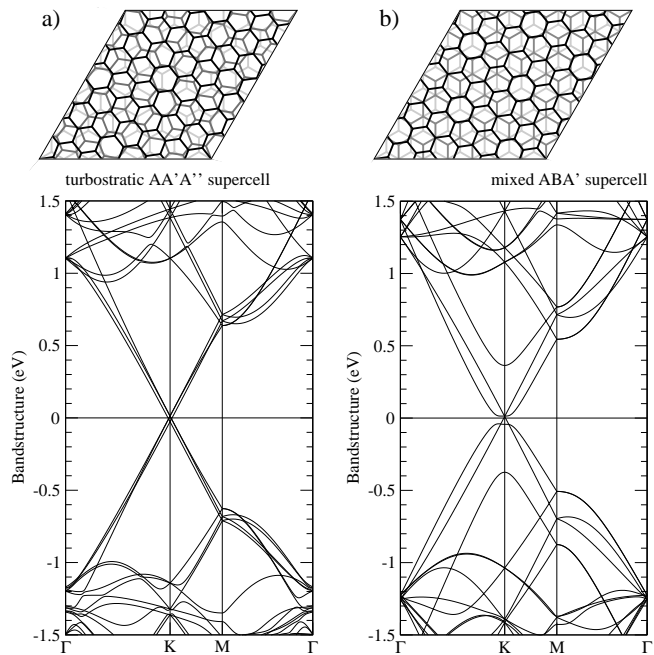,
  width=\linewidth, 
  clip=}
\caption{Representation of the supercells and the correponding electronic 
band structure for trilayer graphites. The turbostratic 
AA'A'' system (43.57$^{\circ}$ and 38.21$^{\circ}$ misorientation angles) 
is shown in (a), and the mixed ABA' structure is shown in (b).
}
\label{f:bsII}
\end{figure}

Turning to the trilayer FLG system, we use the interlayer distance
found for bilayer cases, depending of the stacking geometry (AB or
turbostratic). 
The PAO electronic calculations shown here have been checked 
against PW calculations.
On Fig.\ref{f:bsII}, we present the
results for a {purely} AA'A'' turbostratic
$(5,3)$on$(8,\overline{3})$on$(7,0)$ trilayer with two respective
misorientations of 43.57$^{\circ}$ and 38.21$^{\circ}$, and for a
mixed case $(5,3)$on$(\overline{7},0)$on$(7,0)$, made from a bilayer
with AB stacking and a third one misoriented (38.21$^{\circ}$), called
ABA'. The results for the calculation for ABA or ABC stacking can be
found in references~\cite{Latil_PhysRevLett_98_036803_2006,
Aoki_SolidStateComm_142_123_2007}.

The trilayer with a mixed stacking ABA' displays electronic bands
(Fig. \ref{f:bsII}b) similar to a superposition of a AB bilayer
(Fig. \ref{f:bsI}a) and a single-layer graphene (linear
dispersion). The band crossing of the graphene-like bands
lies $12$~meV above the Fermi level, indicating a small charge transfer of
\mbox{$\sim10^{11}$~e cm$^{-2}$}. 
The other (quadratic) bands present a gap opening 
that is consistent with the behavior found for AB bilayer graphene
under an electric field \cite{Aoki_SolidStateComm_142_123_2007}. 
We
have checked that the same 
systems but with the A' layer away from its
equilibrium position presents a vanishing small charge transfer,
i.e. a progressive closure of the 'gap' of the quadratic bands and a
Dirac point of the linear bands moving towards the Fermi level.
Interestingly, the turbostratic AA'A'' trilayer follows the trends
observed for the bilayer and displays linearly dispersive bands and
massless Fermion behavior. The (doubly degenerate) bands associated
with the external layers are slightly shifted upward ($\sim$12 meV) 
while the bands related to the
central layer are slightly shifted downward($\sim$25 meV).
This is also the signature of a small electron transfer, 
from the external layers to the central one, consistent with 
the previous ABA' case. 
We have also checked that the deviation from the linear
dispersion of the $\pi$-bands for $|E| > 0.5$ {eV} 
is mainly due to the
trigonal warping effect of graphene and not to an
interlayer effect.

The results presented above demonstrate that turbostratic stacking of
graphene sheets leads to a linear dispersion of the electronic levels
similar to the behavior of SLG.  Furthermore, the mixed stacking case
shows a superposition of the two types of carriers, i.e. a massless
Fermion and a {massive,} normal electronic behavior.  Consequently,
direct experimental evidence of Dirac Fermion behavior can no longer
be considered as a discriminating property between single and
multilayered systems.  In particular, the number of $\pi$ bands in
Angle resolved photoemission spectroscopy (ARPES)
measurement~\cite{Ohta_Bostwick}
is not a definitive evidence of the number of layers but should be
cross-checked with other experimental techniques.

The absence of a net effect of interlayer interaction on the
electronic dispersion of turbostratic FLG also suggests that the
$\sqrt{B}$ dependence of the Landau levels will also occur for
multilayer structures and that the abnormal QHE cannot be
excluded. Remarkably, this consequence of our study is supported by
recent experimental data.  First, $\sqrt{B}$ Landau level separation
have been observed in 3-5 graphene
layers~\cite{Sadowski_PhysRevLett_97_266405_2006}.  Second, Dirac
Fermion behavior, integer QHE~\cite{Kempa_SolidStateComm_138_118_2006}
and infrared probe anomalous magnetotransport have been also reported
in a Highly Ordered Pyrolytic Graphite (HOPG) bulk 3D
sample~\cite{Li_and_Li}.  Third, the unconventional
QHE has been reported on bilayer
graphene~\cite{Novoselov_NaturePhys_2_177_2006}, challenging the
common interpretation in terms of AB stacking.  Finally, other
studies~\cite{Lukyanchuk_Zhou} relate the coexistence of {two} types
of carriers (Dirac Fermion and normal electron) in HOPG bulk graphite,
consistent with the findings presented herein.

Recently, the observation of a single 2D band ($\sim$ 2700 cm$^{-1}$)
in Raman spectroscopy has been proposed as an experimental tool for
the determination of the number of layers in FLG samples
\cite{Gupta_Nanolett_6_2667_2006, Ferrari_PhysRevLett_97_187401_2006,
  Graf_Nanolett_7_238_2007}. For two-layer (and thicker) films, the
interlayer interaction induces a splitting of the electronic
bands. This means that in a double resonance Raman process, the
resulting effect is a splitting of this $2D$
peak~\cite{Ferrari_PhysRevLett_97_187401_2006,
  Graf_Nanolett_7_238_2007}.  Our results show that the effect of the
interlayer interaction on the splitting of the electronic bands
crucially depends on the relative orientation of the layers.  For
instance, since a turbostratic bilayer {possesses} degenerate
electronic bands near the Fermi level, the $2D$ mode of this structure
gives rise to a single peak that is indistinguishable from the
signature of SLG systems.  Similarly, a trilayer turbostratic
structure will also present a single $2D$ band.  Indeed, as we have
discussed above, the splitting of the electronic bands in that case is
due to a charge transfer and no evidence of covalent mixing has been
found.  The $2D$ {peak} of the mixed ABA' trilayer will present three
main features. Two of them are associated with the AB-like bands, the
third one with the A' layer, identical to the SLG case.

Interestingly, the splitting of the $2D$ band of graphitic materials
has been found to be a good criteria of crystallinity perpendicular to
the basal planes. For instance, in
Ref.~\onlinecite{Lespade_Carbon_22_375_1984}, the authors have
correlated the interlayer distance in bulk graphite with the doubling
of the $2D$ band and they found that for $d > 0.338$~nm a single $2D$
band is observed where a double $2D$ band is observed for $d
<0.338$~nm. From these findings and the results presented here, we can
propose that HOPG graphite with disorder in the staking (and large
$d$) will present a single $2D$ line whereas Bernal graphite with
(partial) AB order will present a doubling of the $2D$ line. 
>From this discussion, it is clear that a unequivocal determination
of the number of layers is not possible from Raman spectroscopy alone.

\begin{figure}[b]
\epsfig{file=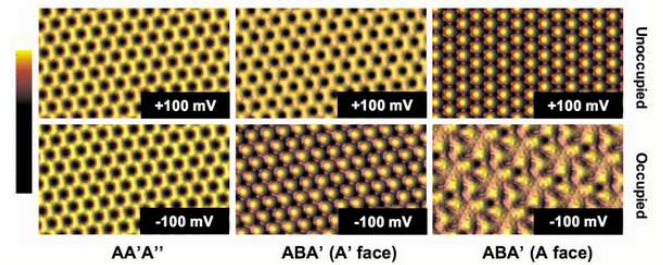, width=\linewidth, clip=}
\caption{
(color online)
>From left to right: constant current STM image simulation of 
turbostratic AA'A'' and mixed ABA' systems. 
(Un)occupied states are images at (positive) negative tip potential. 
The color scale spans the corrugation of each system: 
1.6 (1.4) {\AA}, 
2.3 (2.4) {\AA},
 and 1.8 (2.1) {\AA} 
for each system at positive (negative) bias.
}
\label{f:stm}
\end{figure}

We now analyse the STM images of the trilayer systems studied above.
Recent experimental data on mono- and bilayer graphite on SiC
substrate show a $\sqrt{3}\times\sqrt{3}$ modulation of the STM
intensity. This effect has been associated with the substrate
reconstruction~\cite{Mallet_condmat0702406,
  Rutter_Science_317_219_2007, Varchon_condmat0702311} rather than to
the relative arrangement of the graphitic layers.  Here, we have
computed the STM images of trilayer systems without the substrate in
order to discriminate the intrinsic response of FLG from the effect of
interaction with the substrate.  This approach is made possible by the
prerequisite of Tersoff-Hamann theory since the image formation is
governerd by integration of local density of states, and does not
require a closed circuit to simulate the current\cite{haman}. Here we
made use of the \textsc{BSKAN} package implementation of Tersoff-Hamann
\cite{bskan,notestm}.  Constant current images for systems represented
in Fig.~\ref{f:bsII} are shown in Fig.~\ref{f:stm} for positive and
negative tip polarities.  For system ABA' the two external faces are
not equivalent and the corresponding images are given for both faces.
At low bias (100 meV was used here), the images reflect the property
of the states close to the Fermi level. For system AA'A''', both
positive and negative bias images look the same, with all the atoms
being imaged, just as in SLG.  The situation is quite different for
ABA' systems. The image for occupied states of the A' face shows
bright spots on every other atom, while a complete honeycomb lattice
is imaged for unoccupied states.  The situation is even more
complicated for images computed on the A face of the ABA'
system. While every other atom is now revealed at tip bias
corresponding to unoccupied states, the occupied state image clearly
shows a small-scale Moir\'e-like pattern. It is important to note that
the images have been simulated for zero temperature conditions: given
the very small energy difference between the states responsibles for
the images at low bias, the room temperature image will most
probably look like an average of the images reproduced here.

In summary, we have performed {\it ab initio} calculations of the
electronic band structures of misoriented (turbostratic) FLG. We
surprisingly found that, as a consequence of the incommensurability of
the layers, such multilayer systems possess a massless Fermion
carriers property. Since that property was up to now considered as a
unique signature of single layer graphene, the present findings
challenge the interpretation of the experimental data on FLG.  They
also release the restriction to SLG sample to build devices based on
massless fermion physics.

Calculations were performed at the Namur Interuniversity Scientific
Computing Facility (I-SCF), a common project between the FRS-FNRS and
the University of Namur.  S.L. and L.H. acknowledge the financial
support from the Belgian FNRS. V.M. acknowledges the Division of
Materials Sciences, US Department of Energy, under Contract
No. DEAAC05-00OR22725 with UT Battelle, LLC at ORNL.
%
%
\bibliographystyle{prsty}

\begin{thebibliography}{10}
\bibitem{Note:3layers}
It is impossible to create a totally 
misoriented trilayer structure based on the
  (2,1) supercell, that is still compatible with 
the translationnal symmetry requirements.
\end{thebibliography}

\end{document}